\documentclass[12pt]{iopart}

\usepackage{graphicx}

\begin{document}

\title[Strong Goos-H\"{a}nchen effect of a Gaussian light beam ]
{Direct calculation of the strong Goos-H\"{a}nchen effect of a Gaussian light beam due to the excitation of surface plasmon polaritons in the Otto configuration}

\author{Sangbum Kim$^{1}$ and Kihong Kim$^{1,2}$}

\address{$^1$ Department of Energy Systems
Research and Department of Physics, Ajou University, Suwon 16499, Korea}
\address{$^2$ School of Physics, Korea Institute for Advanced Study, Seoul 02455, Korea}

\ead{khkim@ajou.ac.kr}

\begin{abstract}
We study theoretically the influence of the surface plasmon
excitation on the Goos-H\"{a}nchen lateral shift of a $p$-polarized
Gaussian beam incident obliquely on a dielectric-metal bilayer in
the Otto configuration. We find that the lateral shift depends
sensitively on the thickness of the metal layer and the width of the
incident beam, as well as on the incident angle. Near the incident
angle at which surface plasmons are excited, the lateral shift
changes from large negative values to large positive values as the
thickness of the metal layer increases through a critical value.
For wide incident beams, the maximal forward and backward lateral shifts can
be as large as several hundred times of the wavelength.
As the width of the incident Gaussian beam decreases,
the magnitude of the lateral shift decreases rapidly, but the ratio
of the width of the reflected beam to that of the incident beam,
which measures the degree of the deformation of the reflected beam
profile, increases. In all cases
considered, we find that the reflected beam is split into two parts.
We also find that the lateral shift of the transmitted beam is
always positive and very weak.
\end{abstract}

\maketitle


\section{Introduction}

A light beam deviates from the path expected from geometrical optics
when it is totally reflected at the interface between two different
media. The reflected beam is displaced laterally along the interface,
which is called the Goos-H\"{a}nchen (GH) shift. This phenomenon
has been predicted a long time ago and measured experimentally by Goos and
H\"{a}nchen for the first time \cite{Goos1,Goos2,Goos3}. Artmann
has derived an analytical formula for the GH shift for
incident plane waves \cite{Artmann}. The GH effect occurs in many diverse areas such as
acoustics, optics, plasma physics and condensed matter physics \cite{Lotsch,Puri}.

Earlier works have treated the GH shift in multilayered structures in the
Otto or Kretschmann configuration \cite{Shah1,Shah2,Shah3}. Tamir and Bertoni
performed a detailed analysis of the electromagnetic field distribution
in a leaky-wave structure upon which a Gaussian beam is incident \cite{Tamir}. It
has been demonstrated that the reflected beam displays either
a forward or a backward beam shift. An approximate analytical
solution has shown that the initial Gaussian beam profile splits into
two. The theory of leaky waves has also been applied to acoustic
beams incident on a liquid-solid interface, with the aim of presenting a unified theory of
the beam shifting effect near the Rayleigh angle \cite{Bertoni}. 

The GH effect of a light beam incident on a dielectric slab
from the air has been studied with an emphasis on the transmitted beam
\cite{Li}. Lakhtakia has pointed out that the GH shift
reverses its direction when $\epsilon < 0$ and $\mu < 0$ in the
optically rarer medium \cite{Lakhtakia}. The enhancement of the GH shift and the control of the
reflected field profile has been achieved by adding a defect or cladding layer to
photonic crystals \cite{He,Wang}. Recently, De Leo {\em et al.} have performed 
an extended study investigating the asymmetric GH effect and derived an expression for the GH shift valid 
in the region where the Artmann formula diverges \cite{Leo1,Leo2,Leo3,Leo4}.

Light waves confined to the surface of a medium and surface
charges oscillating resonantly with the light waves constitute the
surface plasmon polaritons (SPPs). The enhancement of electromagnetic
fields near the surface caused by the excitation of SPPs has generated practical applications in
sensor technology \cite{Homola,Kneipp,Kurihara}. These applications
include thin film probing \cite{Pockrand}, biosensing
\cite{Liedberg} and biological imaging \cite{Okamoto1}. In the Otto
or Kretschmann configuration, the SPPs are excited by attenuated
total internal reflection by enhancing the momentum of the incident
light \cite{Yeatman,Torma}.

The excitation of SPPs in the Otto or Kretschmann configuration
affects the GH shift profoundly. Early results on the influence of
SPPs on the shift of light beams can be found in \cite{Mazur} and
\cite{Kou}. It has been shown that the interaction of leaky waves
with SPPs enhances the GH shift. Results on the excitation of
surface waves in the Otto configuration have been reported by Chen
{\em et al.} \cite{Chen}. Chuang has conducted an analysis of the
behavior of the reflection coefficient for both Otto and Kretschmann
configurations \cite{Chuang}. The zeros and poles of the reflection
coefficient move around the complex plane with the change of
parameters, such as the beam width, the wavelength, the thickness
and the dielectric constants. Zeller {\em et al.} have shown that
the coupling of an incident wave with the SPP is highly dependent on
the thickness of the dielectric sublayer in both the Kretschmann and
Otto configuration \cite{Zeller1,Zeller2,Zeller3}. Shadrivov {\em et
al.} have studied the GH shift in the Otto configuration with the
metal sublayer substituted by a left-handed metamaterial
\cite{Shadrivov}. A large GH shift with beam splitting was observed,
and the energy transfer between the right- and left-handed materials
was demonstrated by numerical simulations.
There also exist studies to enhance the GH shift using various hybrid structures 
containing sublayers of graphene, MoS$_2$ or cytop \cite{Xiang1,Xiang2,Xiang3}.
Recently, much progress has been made on obtaining a
tunable GH shift in the prism-coupling system, by applying an external voltage
to a graphene sublayer and other heterostructures \cite{Farmani1,Farmani2,Farmani3,Xiang4}. Kim {\em et al.} have
studied the GH shift of incident $p$ waves in the Otto configuration
containing a nonlinear dielectric layer and shown that its magnitude
can be as large as several hundred times of the wavelength at the
incident angles where the SPPs are excited \cite{Kim3}. Furthermore,
they have shown that the sign and the size of the GH shift can
change very sensitively as the nonlinearity parameter varies.

In this paper, we study the strong enhancement of the GH effect for incident Gaussian
beams when SPPs are excited at the
metal-dielectric interface in the Otto configuration. We examine the
influence of varying the thickness of the metal layer and the incident beam width on the GH effect and find out
optimal configurations for maximal forward and backward lateral shifts.

Our theoretical method is based on the invariant imbedding method,
using which we transform the wave equation into a set of invariant
imbedding equations to create an equivalent initial value problem \cite{Kim5,Kim1,Kim2,Kim6}. 
For the simplest case of multilayered structures made of uniform linear media, this method
is equivalent to those based on the Fresnel coefficients. The invariant imbedding method
has been employed to calculate the GH shift for plane waves incident on nonlinear
media \cite{Kim3,Kim4}. It can also be applied to the case of graded media. Here we consider the interaction
of a Gaussian beam with linear media. More details of our model and method
will be presented in the next section.

\section{Generalization of the invariant imbedding method to Gaussian beams}

We assume the layered structure lies in $0 \le z \le L$. A Gaussian
beam with a finite half-width $W$ is incident from the region where
$z>L$ at an angle $\theta_i$.
For a $p$-polarized beam propagating in the $xz$ plane,
the $y$ component of the magnetic field associated with the
incident beam at the $z=L$ plane can be written as
\begin{equation}
{H_y}^{(i)}(x,L) = H_0 \exp \left( -{x^2 \over {W_x}^2} + i k_{x0} x
\right),
\end{equation}
where $W_x$ $( = W / \cos\theta_i)$ is the half-width in the $x$ direction.
The center of the incident
beam is at $x = 0$.
The parameter $k_{x0}$ ($= k_1 \sin\theta_i$) is the $x$ component of the
wave vector corresponding to the incident angle $\theta_i$ and $k_1$ is the wave number in the incident region, which corresponds to the prism. The superscript $(i)$ refers to the incident
beam.

We consider the incident Gaussian beam as a linear combination of plane waves
and write its field as
\begin{equation}
{H_y}^{(i)}(x,L) = {1 \over \sqrt{2\pi}} \int_{-\infty}^{\infty}
\tilde{H}\left(k_x\right) \exp \left( i k_x x  \right) dk_x,
\end{equation}
where the Fourier transform $\tilde{H}\left(k_x\right)$ is given by
\begin{equation}
\tilde{H}(k_x) = \frac{1}{\sqrt{2}}H_0 W_x\exp\left[-\frac{{W_x}^2}{4}\left(
k_x-k_{x0}\right)^2\right].
\end{equation}
The variable $k_x$ can be parameterized as $k_x=k_1\sin\theta$.
We write the reflection and transmission coefficients corresponding to each Fourier component
as $r(k_x)$ and $t(k_x)$ respectively. Then the field profiles for the reflected and
transmitted beams
${H_y}^{(r)} (x,z)$ and ${H_y}^{(t)} (x,z)$ are given by
\begin{eqnarray}
&&{H_y}^{(r)}(x,z) = {1 \over \sqrt{2\pi}} \int_{-\infty}^{\infty}
r(k_x) \tilde{H}(k_x) \exp \left[ i k_x x + i k_z ( z - L )
\right] dk_x ~~(z>L),\nonumber\\
&&{H_y}^{(t)}(x,z) = {1 \over \sqrt{2\pi}} \int_{-\infty}^{\infty}
t(k_x) \tilde{H}(k_x) \exp \left( i k_x x - i k^\prime_z z
\right) dk_x ~~(z<0),
\end{eqnarray}
where $k_z$ ($=k_1\cos\theta$) and ${k_z}^\prime$ are the negative $z$ components of
the wave vector in the incident ($z>L$) and transmitted ($z<0$) regions respectively.

The GH shifts for the reflected $(\Delta_r)$ and
transmitted $(\Delta_t)$ beams, which are also known as the normalized first momenta
of the magnetic field \cite{Shadrivov}, are defined by
\begin{equation}
\Delta_{r} = \frac{\int_{-\infty}^{\infty} x \left\vert {H_y}^{(r)}(x,L) \right\vert^2 dx} {\int_{-\infty}^{\infty} \left\vert {H_y}^{(r)}(x,L) \right\vert^2 dx},~~~\Delta_{t} = \frac{\int_{-\infty}^{\infty} x \left\vert {H_y}^{(t)}(x,0) \right\vert^2 dx} {\int_{-\infty}^{\infty} \left\vert {H_y}^{(t)}(x,0) \right\vert^2 dx}.
\end{equation}
The reflected and transmitted beams will be severely
deformed when the half-width of the incident beam is small. In order to
measure the degree of the deformation of the reflected and transmitted beams,
we calculate the normalized second momenta of the magnetic field defined by
\begin{eqnarray}
&&\beta_{r} = \frac{4 \int_{-\infty}^{\infty} \left( x - \Delta_{r} \right)^2 \left\vert {H_y}^{(r)}(x,L) \right\vert^2 dx }
 {\int_{-\infty}^{\infty} \left\vert {H_y}^{(r)}(x,L) \right\vert^2 dx},\nonumber\\
&& \beta_{t} = \frac{4 \int_{-\infty}^{\infty} \left( x - \Delta_{t} \right)^2 \left\vert {H_y}^{(t)}(x,0) \right\vert^2 dx }
 {\int_{-\infty}^{\infty} \left\vert {H_y}^{(t)}(x,0) \right\vert^2 dx}.
\end{eqnarray}
These expressions are just the first and second moments of a
distribution function and give the average and the variance for a given
statistical distribution function. The first moment can also be
interpreted as the centroid of the area under the distribution
curve. The half-widths of the reflected and
transmitted beams, $W_r$ and $W_t$, are obtained using
\begin{equation}
W_r = \sqrt{\beta_{r}} \cos\theta_i,~~~W_t = \sqrt{\beta_{t}} \cos\theta_i.
\end{equation}

In the expressions given above, the reflection and transmission
coefficients play a crucial role. In order to calculate $r(k_x)$ and
$t(k_x)$ in the wave vector space, we resort to the invariant imbedding
method, which we summarize here briefly.
For a $p$-polarized wave propagating in a nonmagnetic ($\mu
= 1$) layered structure along the $xz$ plane, the complex amplitude of
the magnetic field, $H = H(z)$, satisfies the wave equation
\begin{equation}\label{p-wave_eq}
{d^2 H \over dz^2} - {1 \over \epsilon(z)} {d \epsilon \over dz} {dH
\over dz} + \left[ {k_0}^2 \epsilon(z) - {k_x}^2 \right] H = 0,
\end{equation}
where $k_0$ ($ = \omega / c$) is the vacuum wave number and
$\epsilon(z)$ is the dielectric permittivity.
We consider a plane wave of unit magnitude $\hat{H}(x,z) = H(z)
e^{ik_xx} = e^{ik_z(L-z)+ik_xx}$ incident on the medium lying in $0\le z\le L$ from the region where $z>L$ at an angle
$\theta$. The complex reflection and transmission coefficients $r =
r(L)$ and $t = t(L)$, which we consider as functions of $L$, are defined by the wave functions outside the
medium by
\begin{equation}
\hat{H}(x,z)=\cases{e^{ik_z(L-z)+ik_xx} + r(L) e^{ik_z(z-L)+ik_xx}, &if $z > L$\\
t(L) e^{-i{k_z}^\prime z+ik_xx},&if $z < 0$\\}.
\end{equation}
When $\epsilon$ is equal to $\epsilon_1$ in $z>L$ and $\epsilon_2$ in $z<0$, the wave vector components
$k_x$, $k_z$ and ${k_z}^\prime$ are given by $ k_x= \sqrt{\epsilon_1} k_0 \sin \theta=k_1\sin\theta$, $ k_z=
\sqrt{\epsilon_1} k_0 \cos \theta$ and ${k_z}^\prime = \sqrt{\epsilon_2}
k_0 \cos \theta^\prime$, where $\theta^\prime$ is the angle that outgoing waves
make with the negative $z$ axis.

We can transform the wave equation into a set of differential
equations for the reflection and transmission coefficients using the invariant imbedding method \cite{Kim5}:
\begin{eqnarray}
{1 \over k_z} {dr(l) \over dl} & = & 2i {\epsilon(l) \over \epsilon_1}
r(l) - {i \over 2} \left[ {\epsilon(l) \over \epsilon_1} - 1 \right]
\left[ 1 - {\epsilon_1 \over \epsilon(l)}
\tan^2 \theta \right] \left[ 1 + r(l) \right]^2, \nonumber \\
{1 \over k_z} {dt(l) \over dl} & = & i {\epsilon(l) \over \epsilon_1}
t(l) - {i \over 2} \left[ {\epsilon(l) \over \epsilon_1} - 1 \right]
\left[ 1 - {\epsilon_1 \over \epsilon(l)}
\tan^2 \theta \right] \left[ 1 + r(l) \right] t(l).
\end{eqnarray}
These equations are integrated from $l=0$ to $l=L$ using
the initial conditions given by the Fresnel
formulas
\begin{eqnarray}
&&r(0) = {\epsilon_2 \sqrt{\epsilon_1} \cos \theta - \epsilon_1
\sqrt{\epsilon_2 - \epsilon_1 \sin^2 \theta} \over \epsilon_2
\sqrt{\epsilon_1} \cos \theta + \epsilon_1 \sqrt{\epsilon_2 -
\epsilon_1 \sin^2 \theta}}, \nonumber\\
&&t(0) = {2 \epsilon_2 \sqrt{\epsilon_1} \cos \theta \over
\epsilon_2 \sqrt{\epsilon_1} \cos \theta + \epsilon_1
\sqrt{\epsilon_2 -
\epsilon_1 \sin^2 \theta}}.
\end{eqnarray}
The reflection and transmission coefficients $r(L)$ and $t(L)$
calculated above are employed into the integral in (4),
multiplying each Fourier component.

The GH shift for reflected plane waves is computed and compared with
those of beams, using Artmann's formula
\begin{equation}
\Delta = -{d\phi \over dk_x} = -{\lambda \over 2\pi
\sqrt{\epsilon_1} \cos\theta} {d\phi \over d\theta}
\end{equation}
where $\lambda$ is the wavelength and $\phi$ is the phase of the
reflection coefficient satisfying $r = |r| e^{i \phi}$.

\section{Numerical results and discussion}

\begin{figure}[htbp]
\centering
\includegraphics[width=15cm]{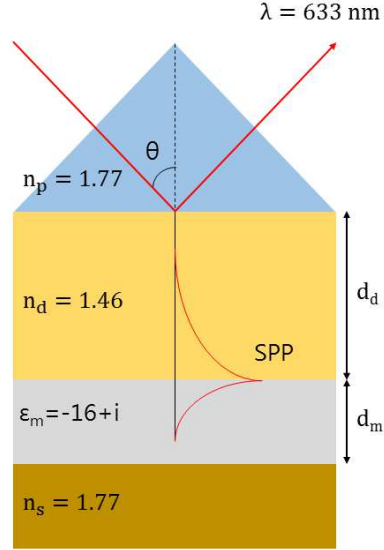}
\caption{\label{fig1} Schematic of the Otto configuration considered in this paper. }
\end{figure}

We assume that a $p$-polarized beam is incident from a prism onto a
dielectric-metal bilayer, which lies on a dielectric substrate, at room temperature. Both
the prism and the substrate are assumed to have the same refractive
index of 1.77 corresponding to sapphire. The refractive index of the dielectric
layer is 1.46 corresponding to fused silica
(${\rm SiO_2}$). Then the critical incident angle $\theta_c$ is equal
to $55.57^\circ$. The vacuum wavelength of the incident wave
$\lambda$ is  633 nm and the dielectric permittivity of the metal
layer $\epsilon_m$ is $-16+i$ corresponding to silver at $\lambda =
633$ nm. In figure~1, we show the schematic of our configuration.
We note that we have a propagating wave in the
substrate in the present geometry.

\begin{figure}[htbp]
\centering
\includegraphics[width=8.5cm]{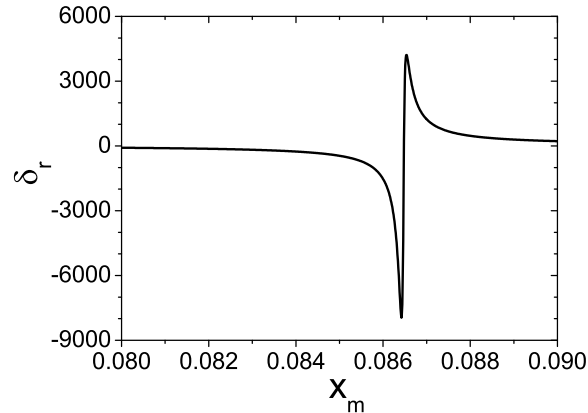}
\caption{\label{fig2} GH shift normalized by the wavelength, $\delta_r = \Delta_r /
\lambda$, of the reflected wave calculated from
Artmann's formula for plane waves when $\theta_i= 62.33^\circ$ plotted versus the thickness of the metal layer
normalized by the wavelength, $x_m = d_m / \lambda$.}
\end{figure}

We first consider the case where plane waves are incident.
In figure~2, we plot the GH shift normalized by the wavelength, $\delta_r = \Delta_r /
\lambda$, of the reflected wave calculated from Artmann's formula for
plane waves versus the thickness of the metal layer normalized by
the wavelength, $x_m = d_m / \lambda$, when $\theta_i$ is fixed to $62.33^\circ$.
As $x_m$ is increased through the
critical value $x_{m,\rm cr}\approx 0.08648$, $\delta_r$ changes
rapidly from large negative values to large positive values. The
value of the largest backward normalized GH shift is about $-7957$
at $x_m\approx 0.08642$, while that of the largest forward
normalized GH shift is about 4214 at $x_m\approx 0.08654$.
We note that the change of $x_m$ from $0.08642$ to $0.08654$ corresponds to the thickness change of just 0.076 nm.
From many numerical calculations, we have found that the maximum GH shifts are
obtained when $\theta_i\approx 62.33^\circ$. Since the values of the GH shift for plane waves are
much larger than those obtained for Gaussian beams and shown in figure~4(a) below, we show them in
separate figures.

Shen {\em et al.} have presented the results of
the calculation and the measurement of the phase shift of the
reflected wave in the Kretschmann configuration as a function of the
incident angle \cite{Shen}. It has been shown that the phase shift
decreases monotonically with $\theta_i$, which corresponds to the
forward GH shift, when the metal (Ag) layer is relatively thin,
while it increases monotonically, which corresponds to the backward
GH shift, when the metal layer is relatively thick. Shen {\em et
al.} have shown that the forward shift changes to the backward shift
when the metal layer thickness changes by only a few nanometers.
We notice that the sign of the GH shift changes from negative to positive
as $d_m$ increases
in our case of the Otto configuration, whereas it changes from positive to negative in Shen {\em et al.}'s
case of the Kretschmann configuration.

\begin{figure}[htbp]
\centering
\includegraphics[width=9cm]{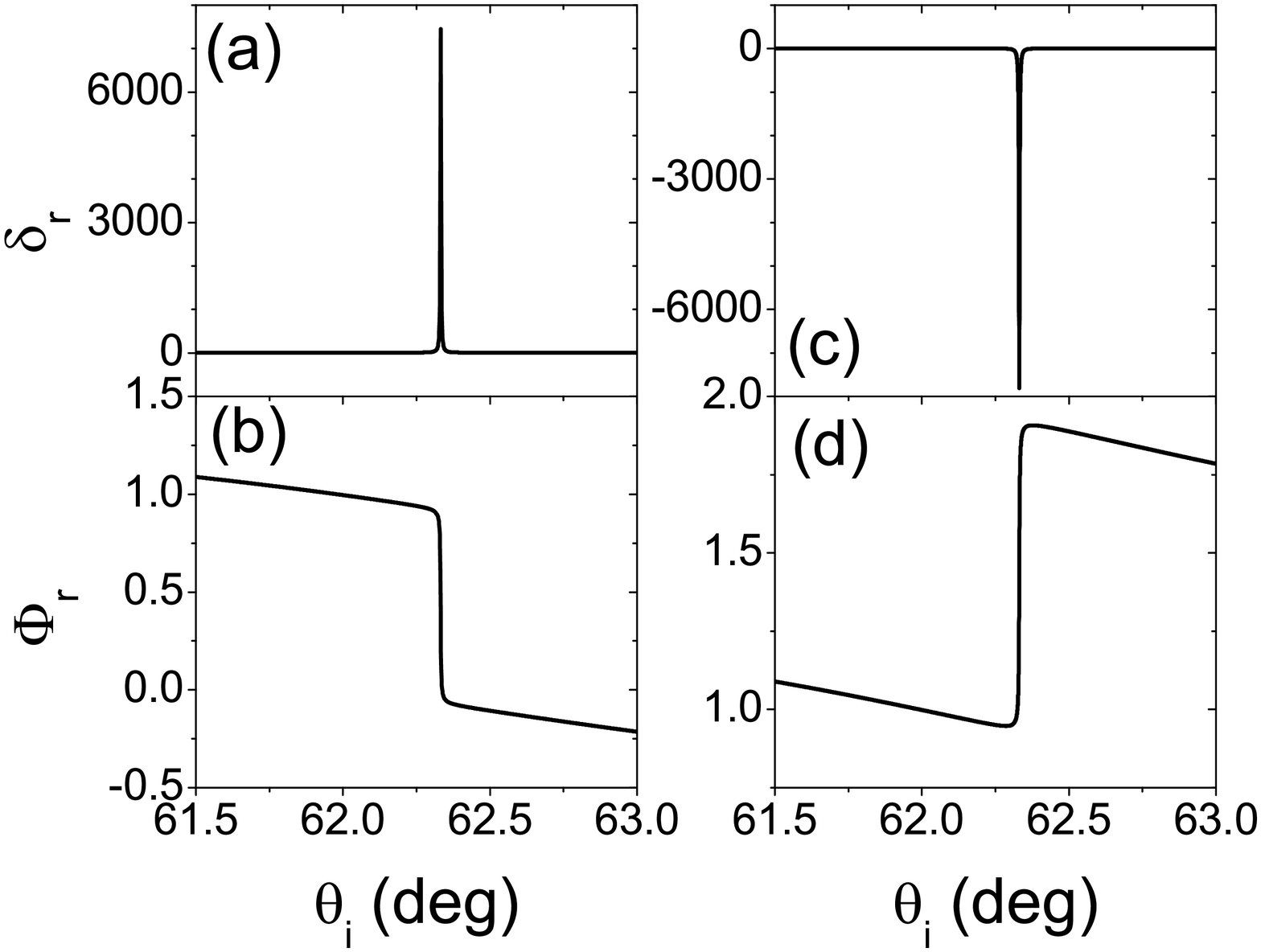}
\caption{\label{fig3} (a) Forward normalized GH shift of the
reflected wave, $\delta_r$, calculated from Artmann's formula for
plane waves when $x_m=0.08657$ and (b) the corresponding phase of
the reflected wave, $\Phi_r$, versus incident angle $\theta_i$. (c)
Backward normalized GH shift of the reflected wave calculated from
Artmann's formula when $x_m=0.08638$ and (d) the corresponding phase
of the reflected wave versus incident angle.}
\end{figure}

In figures~3(a) and 3(b), we show the forward GH shift for the
reflected plane wave and the corresponding phase of the reflected
wave as a function of the incident angle, when $x_m=0.08657$. We
find that $\Phi_r$ changes very rapidly and the GH shift becomes
extremely large near $\theta_i=62.33^\circ$. The maximum $\delta_r$
is about 7450 at $\theta_i=62.332^\circ$. Similarly, in figures~3(c)
and 3(d), we show the backward GH shift for the reflected plane wave
and the corresponding phase of the reflected wave obtained for
$x_m=0.08638$. $\delta_r$ is about $-7814$ at
$\theta_i=62.3311^\circ$ in this case. We find that the maximal
forward and backward GH shifts occur at the same incident angles
where the reflectance takes a minimum value.

\begin{figure}[htbp]
\centering
\includegraphics[width=8.5cm]{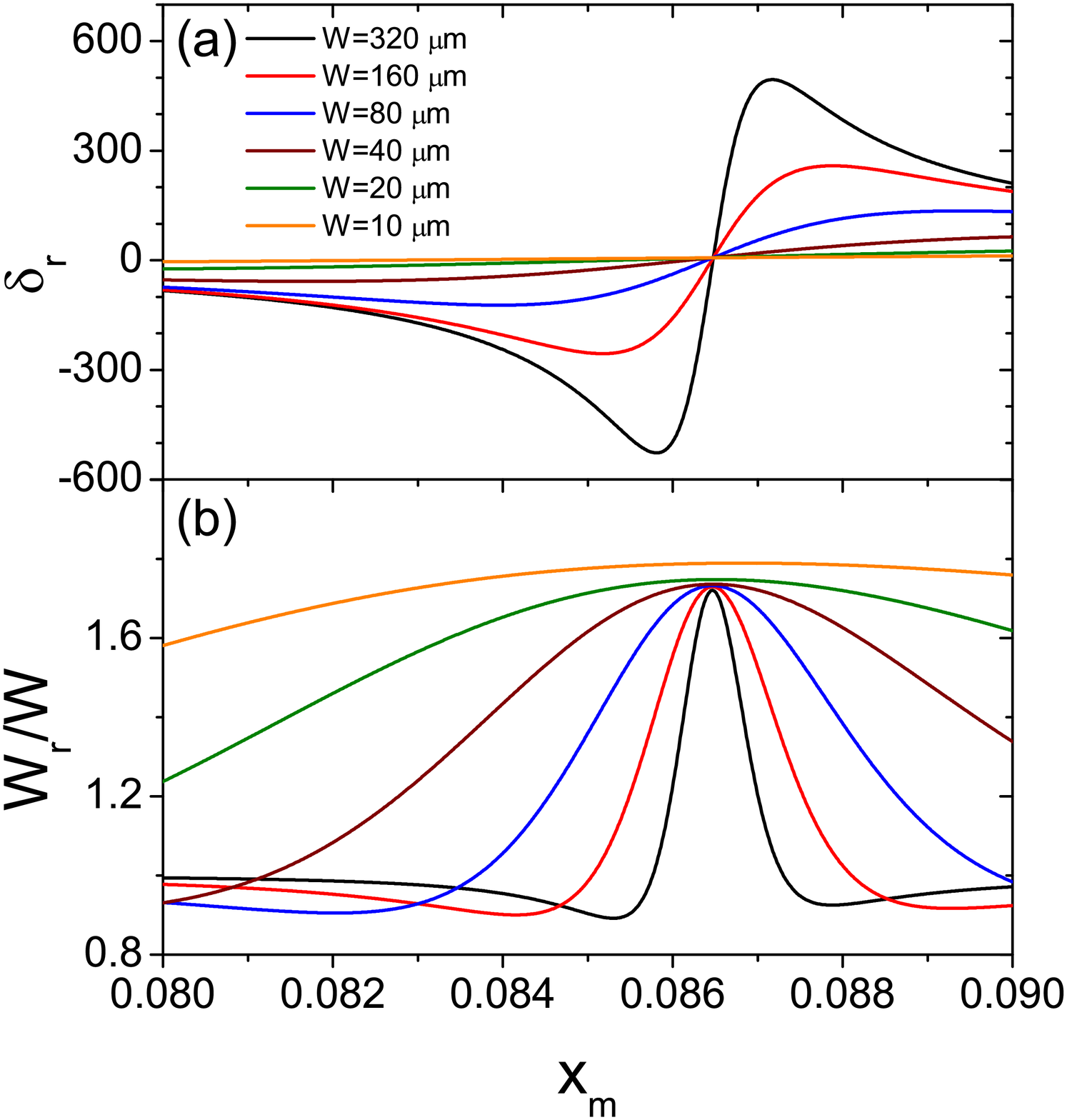}
\caption{\label{fig4} (a) Normalized GH shift of the reflected beam
$\delta_r$ plotted versus $x_m$, for several different values of
the half-width $W$ of the incident Gaussian beam when $\theta_i=
62.33^\circ$. (b) Half-width of the reflected beam $W_r$ normalized
by $W$ plotted versus $x_m$ for the same values of $W$ and
$\theta_i$ as in (a).}
\end{figure}

We now consider the case where Gaussian light beams are incident.
In figure~4(a), we plot the normalized GH shift of the reflected beam
$\delta_r$ versus $x_m$, for several different values of the
half-width $W$ of the incident Gaussian beam. The incident angle is
fixed to $\theta_i= 62.33^\circ$. For the displayed range of $x_m$
$\in [0.08, 0.09]$, $d_m$ varies from 50.64 nm to 56.97 nm. The
thickness of the dielectric layer, $d_d$, is chosen such that $x_d =
d_d / \lambda = 0.5055$, therefore $d_d$ is equal to 319.98 nm. The
specific values of $\theta_i$ and $d_d$ were chosen so as to
maximize the largest value of the GH shift.

We observe in figure~4(a) that the GH shift of the reflected beam
changes its sign from negative to positive as $x_m$ increases
through a critical value $x_{m,\rm cr}$. This value and the
corresponding critical thickness of the metal layer, $d_{m,\rm cr}$,
are found to vary very little as $W$ is decreased from 320 $\mu$m $( \approx 505.53\lambda)$ to
40 $\mu$m $( \approx 63.19\lambda)$. As $W$ is decreased further, $d_{m,\rm cr}$ decreases
noticeably. $x_{m,\rm cr}$ is about
0.0862 and $d_{m,\rm cr}$ is about 54.56 nm when  $W$ is 40 $\mu$m, while $x_{m,\rm cr}$ is about
0.0825 and $d_{m,\rm cr}$ is about 52.22 nm when  $W$ is 10 $\mu$m $(\approx  15.8\lambda)$. The maximal forward and
backward lateral shifts are close to the half-width of the
incident beam. For example, the beam with $W = 505.53\lambda$ is shifted by
$\Delta_r = 494.72\lambda$ in the forward direction and  by $\Delta_r =
-527.2\lambda$ in the backward direction. For $\lambda =
633$ nm, these correspond to 313.2 $\mu$m and $-333.7$ $\mu$m respectively.
Our numerical results are summarized in Table 1. We notice that for $W/\lambda \approx 15.8$, the forward GH shift has no local maximum.

\begin{table}
\centering \caption{\label{tab:tab1} Summary of the numerical
results for varying beam widths $W$. $x_{m,{\rm cr}}$ is the normalized critical thickness of
the metal layer. $x_{m,f}$ and
$x_{m,b}$ are the normalized thicknesses for the maximal forward and
backward GH shifts, while $\delta_{r,f}$ and
$\delta_{r,b}$ are the corresponding maximal forward and backward GH shifts normalized by $\lambda$. $x_{m,{\rm md}}$ is
the normalized thickness corresponding to the maximum
distortion, namely, the maximum $W_r$. $W_{r,{\rm md}}/W$ is the normalized value of the maximum distortion.}

\begin{tabular}{ccccccccc}
\hline
$W$($\mu$m) & $W/\lambda$ & $x_{m,{\rm cr}}$ & $x_{m,f}$ & $\delta_{r,f}$ & $x_{m,b}$ & $\delta_{r,b}$ & $x_{m,{\rm md}}$ & $W_{r,{\rm md}}/W$ \\
\hline
  $\infty$  &     $\infty$  & 0.08648 & 0.08654 & 4214.2 & 0.08642 & $-7956.8$ & -- & -- \\
  320 &     505.53  & 0.08647 & 0.08717 & 494.72 & 0.08581 & $-527.20$ & 0.08647 & 1.72054\\
  160&     252.76  & 0.08646 & 0.08788 & 258.86 & 0.08517 & $-255.52$ & 0.08647 & 1.72957\\
  80&     126.38  & 0.08640 & 0.08939 & 135.19 & 0.08396 & $-122.74$ & 0.08647 & 1.73259\\
  40&     63.19  & 0.08617 & 0.09286 & 72.20 & 0.08173 & $-57.48$ & 0.08648 & 1.73621\\
  20&     31.6  & 0.08530 & 0.10287 & 40.69 & 0.07790 & $-25.45$ & 0.08652 & 1.74811\\
  10&     15.8  & 0.08246 & -- & -- & 0.07196 & $-10.10$ & 0.08683 & 1.78945\\
\hline
\end{tabular}
\end{table}


Our result can be understood from the theory of the GH shift in multilayered structures,
where the sign of the GH shift is determined by
the competition between intrinsic damping and radiative damping \cite{Liu,Okamoto2}. As the
thickness of the dielectric or metal layer is varied, the GH shift becomes
positive if the intrinsic damping (absorption loss in the lossy layer) is
less than the radiative damping (leakage from the dielectric layer to the prism), and
negative otherwise. Our result is consistent with this argument, since the
intrinsic damping is inversely proportional to the thickness of
the metal layer \cite{Okamoto2}. Thus the reduction of $d_m$ enhances the intrinsic damping,
resulting in the negative GH shift in figure~4(a).

Zeller {\em et al.} have also performed a theoretical analysis of
the reflection coefficient associated with $s$-polarized SPPs in the
Otto configuration containing a layer of a negative-index
metamaterial instead of a metal layer \cite{Zeller3}. They have
found that the sign of the GH shift changes from negative to
positive as the thickness of the dielectric layer, $d_d$, increases.

In figure~4(b), we plot the half-width of the reflected beam $W_r$
normalized by $W$ versus $x_m$. This value takes a maximum at $x_m
\approx 0.08647$ for $W =  320$, 160 and 80 $\mu$m. The thickness for
the maximum $W_r/W$ increases slightly to $x_m = 0.08683$ for $W =
10$ $\mu$m. Thus the maximum distortion of the beam occurs near the
critical thickness of the metal layer, but not at the exactly same
thickness. We observe that $W_r/W$ is larger and the overall curve
is broader when the GH shift is smaller and when the incident beam
is narrower.

\begin{figure}[htbp]
\centering
\includegraphics[width=8.5cm]{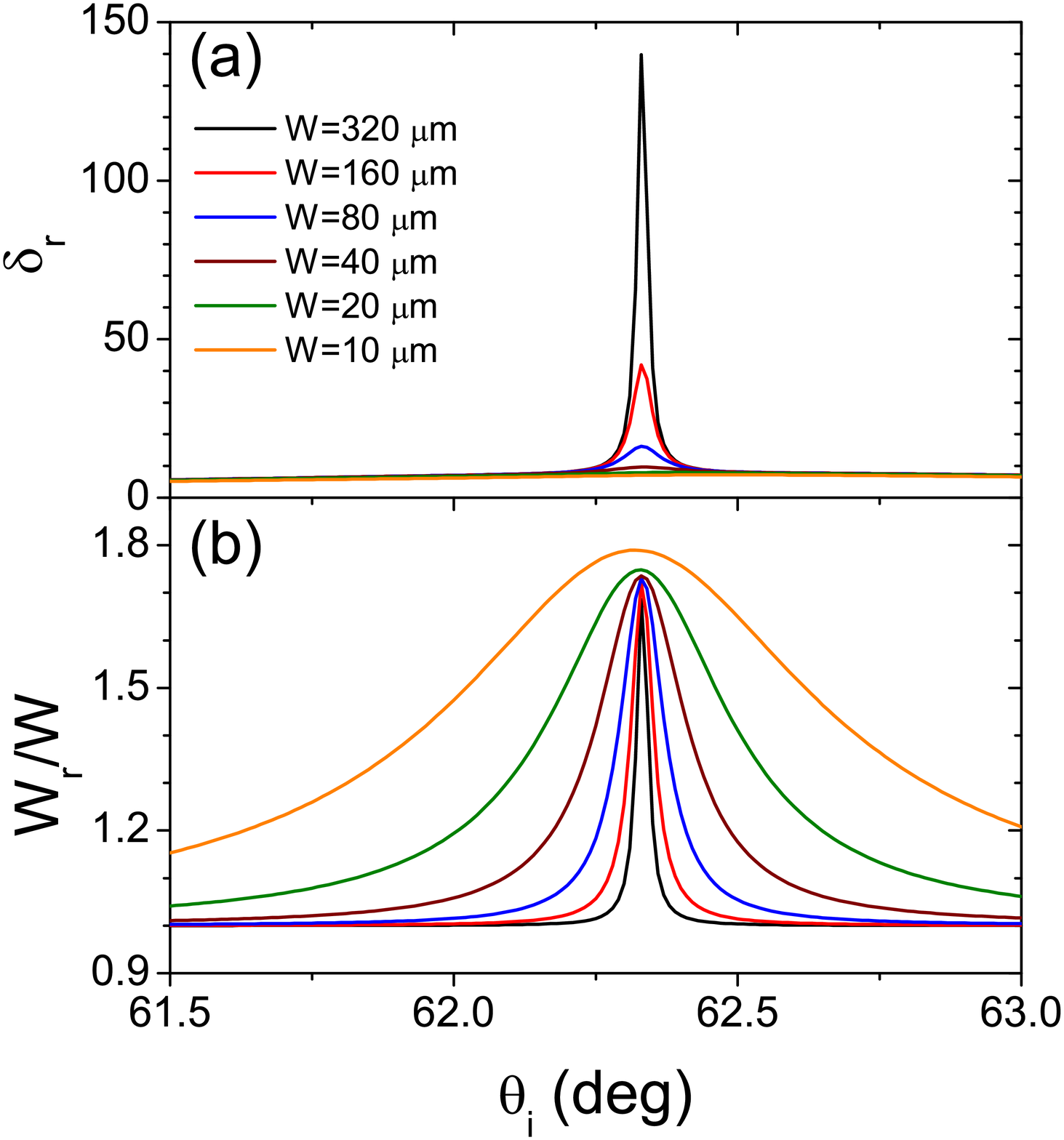}
\caption{\label{fig5} (a) Forward normalized GH shift of the
reflected beam versus incident angle for several different values of
$W$ when $x_m=0.08657$. (b) Normalized half-width of the reflected beam versus incident
angle for the same values of $x_m$ and $W$ as in (a).}
\end{figure}

In figure~5(a), we plot $\delta_r$ as a function of the incident angle
for several values of $W$ when $x_d$ is 0.5055 and $x_m$ is 0.08657.
We find that the maximum value of the {\em forward} lateral shift,
which is about 140 times of the wavelength when $W$ is 320 $\mu$m
and about 42 times of the wavelength when $W$ is 160 $\mu$m, is an
order of magnitude smaller than the value in the plane-wave
incidence, but is still very large. This value decreases as $W$
decreases. Since each Fourier component in the incident beam
experiences a different phase shift from each other, the reflected
beam, which is the sum of the reflected Fourier components, gets a
much smaller GH shift than the case of the plane-wave incidence. As
the half-width of the Gaussian beam increases, the magnitude of the
lateral shift of the reflected beam increases and approaches the
value for the plane-wave incidence. The reflectance for plane waves
takes a dip at $\theta = 62.33^\circ$ and the maximum of $\delta_r$
occurs at the reflectance dip.

In figure~5(b), we show the normalized half-width of the
reflected beam, $W_r / W$, as a function of the incident
angle for each half-width of the incident Gaussian beam. We find
that the half-width of the reflected beam at the reflectance dip is
substantially larger than that of the incident Gaussian beam. The relative distortion of the
reflected beam profile is larger for narrower beams and does not disappear at $\theta_i\approx 62.33^\circ$ as the beam width
$W$ increases.

\begin{figure}[htbp]
\centering
\includegraphics[width=8.5cm]{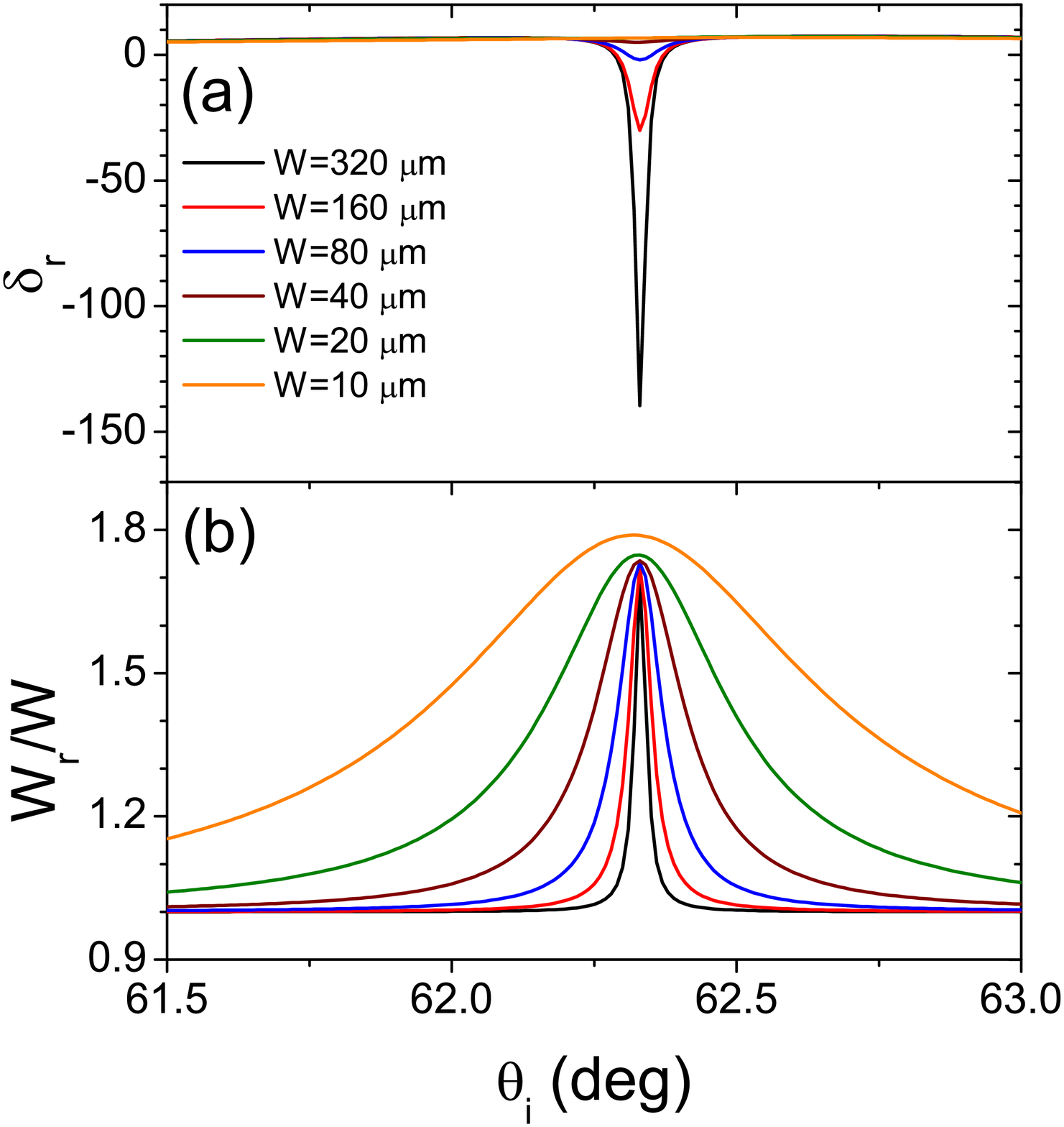}
\caption{\label{fig6} (a) Backward normalized GH shift of the
reflected beam versus incident angle for several different values of
$W$ when $x_m=0.08638$. (b) Normalized half-width of the reflected beam versus incident
angle for the same values of $x_m$ and $W$ as in (a).}
\end{figure}

In figure~6(a), we plot $\delta_r$ as a function of the incident angle for
different values of $W$, when $x_d$ is 0.5055 and $x_m$ is 0.08638. We find that the GH shift at the reflectance dip is {\em backward}
with negative values of $\delta_r$ when $W$ is larger than 40 $\mu$m
in the present case. The maximum value of $\vert\delta_r\vert$ is about 140 when $W$ is 320 $\mu$m
and about 30 when $W$ is 160 $\mu$m.
Interestingly, for $W \leq 40$ $\mu$m, the
value of $\delta_r$ at the reflectance dip becomes positive. The
critical thicknesses in figure~4(a) for narrow beams with $W = 40$,
20 and 10 $\mu$m are $x_m = 0.08617$, 0.08530 and 0.08246 respectively.
Thus in the calculation with $x_m = 0.08638$ presented here,
narrow beams have positive GH shifts.
In figure~6(b), we show the normalized half-width of the reflected beam
$W_r / W$ as a function of the incident angle for each
value of $W$. General characteristics are similar to the forward shift case shown in figure~5(b).

\begin{figure}[htbp]
\centering
\includegraphics[width=8.5cm]{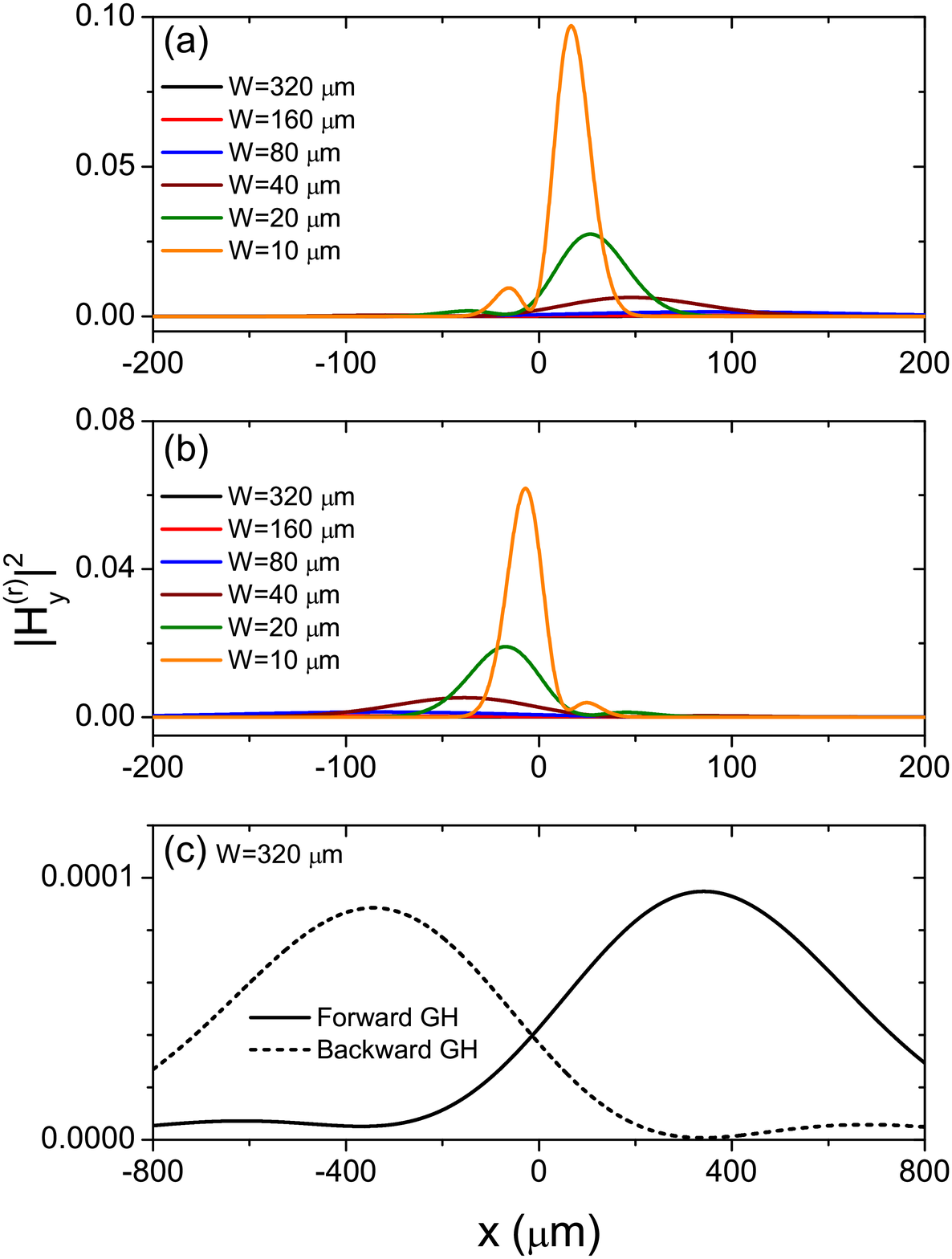}
\caption{\label{fig7} (a) Magnetic field intensity distribution associated with the reflected beam
plotted versus $x$ for several different values of $W$, when $x_m=x_{m,f}$ and $\theta_i= 62.33^\circ$.
The GH shift is positive in this case.  (b) Magnetic field intensity distribution associated with the reflected beam
plotted versus $x$ for several different values of $W$, when $x_m=x_{m,b}$ and $\theta_i= 62.33^\circ$.
The GH shift is negative in this case. (c) Comparison of the magnetic
field profiles associated with the reflected beams corresponding to the backward ($x_m=0.08581$) and forward ($x_m=0.08717$) GH shifts, when $W = 320$ $\mu$m and $\theta_i= 62.33^\circ$.}
\end{figure}

In figures~7(a) and 7(b), we plot the magnetic field intensity distribution associated with the reflected beam at the
reflectance dip ($\theta_i=62.33^\circ$) for several different values of $W$, when $x_m=x_{m,f}$ and $x_m=x_{m,b}$ respectively.
In all cases, we observe clearly that the
reflected beam is split into two parts.
In the case of the plane-wave incidence, the reflectance is very low. This should apply to other Fourier
components as well, so the intensity of the reflected beam is substantially weaker than
that of the incident beam.

In figure~7(c), we compare the magnetic
field profiles associated with the reflected beams corresponding to the backward ($x_m=x_{m,b}$) and forward ($x_m=x_{m,f}$) GH shifts,
when $W = 320$ $\mu$m and $\theta_i= 62.33^\circ$.
We observe that the left peak is larger than the right peak in the case
of the backward GH shift, while it is vice versa in the case of the positive GH shift.
The maximum forward GH shift for the beam with $W = 320$ $\mu$m is about 495 times of the wavelength
($\sim 313$ $\mu$m), which is sufficiently large to be observable in figure~7.

%

We mention briefly on the GH shift of the
transmitted beam, $\delta_t$ ($ = \Delta_t / \lambda$), as a
function of the incident angle for several values of $W$. We find
that $\delta_t$ is positive in all cases and depends very weakly on
$W$. The maximum value of $\Delta_t$ is about seven times of the
wavelength ($\sim 4.4$ $\mu$m). For the normalized
half-width of the transmitted beam $W_t/ W$, we find that the value
at the reflectance dip is slightly larger than 1. We also find that there
is no splitting in the magnetic field intensity distribution of the
transmitted beam. The spatial profiles of the transmitted beam look
similar to those of the incident Gaussian beam.
If we employ a substrate with its
refractive index less than 1.5676, we obtain an evanescent
transmitted beam. Then there will be another SPP excited at the
metal-substrate interface, enhancing the GH shift even further.

There have been several experiments showing both positive and
negative GH shifts. In particular, there exist two experiments that have used experimental
parameters similar to ours \cite{Liu,Yin}. In \cite{Shen}, the totally reflected $p$-polarized wave is heterodyned with the $s$-polarized wave from the same laser source. In \cite{Yin}, a position-sensitive detector produces a signal that is proportional to the difference of the lateral displacements of $s$- and $p$-polarized waves. Since an $s$ wave does not excite SPPs, the obtained signal gives the GH shift of the $p$-polarized wave in both cases.
Following similar schemes, it will
not be difficult to set up an experiment in the Otto configuration
to test our results.


\section{Conclusion}

In this paper, we have studied the influence of the surface plasmon
excitation on the GH effect theoretically, when a Gaussian beam of a
finite width is incident on a dielectric-metal bilayer in the Otto
configuration. Using the invariant imbedding method, we have
calculated the GH lateral shifts of the reflected and transmitted
beams and the widths of these beams relative to that of the incident
beam. We have found that the lateral shift of the reflected beam
depends very sensitively on the thickness of the metal layer and the
width of the incident beam. Close to the incident angle at which
surface plasmons are excited, the lateral shift of the reflected
beam has been found to change from large negative values to large
positive values as the thickness of the metal layer increases
through a critical value. The maximal forward and
backward lateral shifts can be as large as the half-width of the
incident beam. As the width of the
incident Gaussian beam decreases, we have found that the size of the
lateral shift decreases rapidly, but the relative deformation of the
reflected beam increases. In all cases studied, we have
found that the reflected beam is split into two parts. Finally, we
have found that the lateral shift of the transmitted beam is always
positive and very weak. The lateral shift of the reflected beam
studied here is large enough to be considered for application to
optical devices.

\ack
This work has been supported by the National Research Foundation of Korea Grant (NRF-2018R1D1A1B07042629) funded by the Korean Government.

\end{document}